\documentclass[pre,tighten]{revtex4}

\usepackage{amssymb,amsmath}

\usepackage{graphicx}

\begin{document}
\title{Hydrodynamics for inelastic Maxwell mixtures: Some applications}
\author{Vicente Garz\'{o}}
\email[E-mail: ]{vicenteg@unex.es}
\affiliation{Departamento de F\'{\i}sica, Universidad de Extremadura, E-06071
Badajoz, Spain}

\author{Jos\'e Mar\'{\i}a Montanero}
\email[E-mail: ]{jmm@unex.es}
\affiliation{Departamento de Electr\'onica e Ingenier\'{\i}a Electromec\'anica, Universidad de Extremadura, E-06071 Badajoz, Spain}

\pacs{ 05.20.Dd, 45.70.Mg, 51.10.+y, 47.50.+d}
\date{\today}

\begin{abstract}
Hydrodynamic equations for a binary mixture of inelastic Maxwell models described by the Boltzmann equation are derived. The Navier-Stokes transport coefficients are {\em exactly} obtained by solving the Boltzmann equation from the Chapman-Enskog method for states close to the (local) homogeneous cooling state (HCS). The knowledge of the transport coefficients allows one to analyze two different problems. First, we solve the linearized hydrodynamic equations around the homogeneous (cooling) state and identify the conditions for stability as functions of the wave vector, the dissipation, and the parameters of the mixture. As happens for monocomponent systems, the analysis shows that the HCS is unstable to long enough wavelength perturbation. As a second problem, we explore the validity of Onsager's reciprocal relations of a granular binary mixture. As expected, since a granular system is not time reversal invariant, Onsager's reciprocal relations do not apply for inelastic collisions. The results show that the absence of the Gibbs state (non-Maxwellian behavior of the velocity distribution functions describing the HCS), the collisional cooling, and the occurrence of different kinetic temperatures for both species (breakdown of energy equipartition) are responsible for a violation of Onsager's relations.
\end{abstract}

\maketitle

\section{Introduction}
\label{sec1}

Granular fluids are usually modeled as systems composed by smooth hard spheres with inelastic collisions. The collisions are specified in terms of the change in relative velocity at contact but with a decrease of the magnitude of the normal component measured by a (positive) coefficient of restitution smaller than or equal to 1. For this interaction model, the Boltzmann equation (BE) has been conveniently modified to account for the inelasticity of binary collisions and the transport coefficients have been determined for a single gas \cite{BDKS98} as well as for multicomponent systems \cite{GD02}. However, the evaluation of those coefficients for inelastic hard spheres (IHS) is quite involved since they are given in terms of the solutions of linearized integral equations which can be approximately solved by considering the leading order in a Sonine polynomial expansion of the velocity distribution function. For this reason, other interaction models that simplify the mathematical structure of the Boltzmann collision integrals for IHS have been considered in the past few years. As for ordinary gases, the BE for inelastic Maxwell models (IMM) has been also introduced \cite{IMM}. The IMM share with elastic Maxwell molecules the property that the collision rate is velocity independent but, on the other hand, their collision rules are the same as for IHS. Although these IMM do not correspond to any microscopic interaction potential, the cost of sacrificing physical realism is in part compensated by the amount of exact analytical results that can be derived from this simple model.

Most of the studies performed for IMM have been devoted to homogeneous states, especially in the analysis of high-energy asymptotics of the velocity distributions. On the other hand, much less is known on the dependence of the transport coefficients on dissipation, especially for mixtures. In the case of a binary mixture subjected to simple shear flow, the relevant rheological properties of the system (shear and normal stresses) have been exactly obtained \cite{G03}. These results show an excellent agreement with those obtained analytically for IHS in the first Sonine approximation and by means of Monte Carlo simulations \cite{MG02}. More recently, the BE for inelastic Maxwell mixtures has been solved \cite{GA04} from the Chapman-Enskog method for states close to the (local) homogeneous cooling state. Explicit expressions of the Navier-Stokes transport coefficients of IMM in $d$ dimensions have been derived in terms of the coefficients of restitution and the ratios of mass, concentration, and particle sizes. Comparison with known results for IHS \cite{GD02} shows a reasonably good agreement, especially for the transport coefficients associated with the mass flux.

The knowledge of the transport coefficients of IMM for mixtures allows quantitative application of the nonlinear hydrodynamic equations to a number of interesting problems. Here, we consider two different applications. First, we study small perturbations of a spatially homogeneous state and determine the dispersion relations for the hydrodynamic modes. This analysis allow us to identify the conditions for stability as functions of the wave vector, the dissipation, and the parameters of the mixture. As in the monocomponent case \cite{BDKS98}, linear stability analysis shows two transversal (shear) modes and a longitudinal (heat) mode to be unstable with respect to long wavelength excitations. As a second application, we explore the validity of Onsager's reciprocal relations \cite{GM84} among the different transport coefficients associated with the mass and heat fluxes. For ordinary gases these relations are a consequence of time reversal invariance of the equations of motion of the individual particles. Since a granular fluid is inherently time irreversible (there is a irreversible loss of kinetic energy in collisions), violation of Onsager's relations is expected with increasing dissipation. The interesting point here is to assess the influence of inelasticity on the failure of these reciprocal relations as well as to identify the origin of such a violation.

The plan of the paper is as follows. In Section \ref{sec2}, we give a brief summary of the Boltzmann equation for IMM and their corresponding balance hydrodynamic equations. The form of the Navier-Stokes hydrodynamic equations for the mixture is given in Sec.\ \ref{sec3}. Sections \ref{sec4} and \ref{sec5} contain to the main results of this paper. In Sec.\ \ref{sec4} we perform a stability analysis of the linearized hydrodynamic equations while Section \ref{sec5} deals with the study of Onsager's relations in granular systems. We close the paper with some concluding remarks in Sec.\ \ref{sec6}.

\section{Inelastic Maxwell models for a granular mixture}
\label{sec2}

Let us consider a mixture of inelastic Maxwell gases at low density \cite{IMM}. At this level of description, all the relevant information on the system is given through the velocity distribution functions $f_r({\bf r}, {\bf v};t)$ ($r=1,2, \ldots$) of each species, which obey the following set of nonlinear Boltzmann kinetic equations \cite{G03}
\begin{equation}
\label{2.1}
\left(\partial_t+{\bf v}\cdot \nabla \right)f_{r}
({\bf r},{\bf v};t)
=\sum_{s}J_{rs}\left[{\bf v}|f_{r}(t),f_{s}(t)\right] \;,
\end{equation}
where  the Boltzmann collision operator $J_{rs}\left[{\bf v}|f_{r},f_{s}\right]$ is
\begin{eqnarray}
J_{rs}\left[{\bf v}_{1}|f_{r},f_{s}\right] &=&\frac{\omega_{rs}({\bf r},t;\alpha_{rs})}{n_s({\bf r},t)\Omega_d}
\int d{\bf v}_{2}\int d\widehat{\boldsymbol{\sigma }}\nonumber\\
& & \times
\left[ \alpha_{rs}^{-1}f_{r}({\bf r},{\bf v}_{1}',t)f_{s}(
{\bf r},{\bf v}_{2}',t)-f_{r}({\bf r},{\bf v}_{1},t)f_{s}({\bf r},{\bf v}_{2},t)\right]
\;.
\label{2.2}
\end{eqnarray}
Here, $m_r$ is the mass of a particle of species $r$, $n_r$ is the number density of species $r$, $\omega_{rs}\neq \omega_{sr}$ is an effective collision
frequency (to be chosen later) for collisions  of type $r$-$s$,  $\Omega_d=2\pi^{d/2}/\Gamma(d/2)$ is the total solid angle in $d$ dimensions, and $\alpha_{rs}=\alpha_{sr}\leq 1$ refers to the constant coefficient of restitution for collisions between particles of species $r$ and $s$. In
addition, the primes on the velocities denote the initial values $\{{\bf v}_{1}^{\prime},
{\bf v}_{2}^{\prime}\}$ that lead to $\{{\bf v}_{1},{\bf v}_{2}\}$
following a binary collision:
\begin{equation}
\label{2.3}
{\bf v}_{1}^{\prime }={\bf v}_{1}-\mu_{sr}\left( 1+\alpha_{rs}
^{-1}\right)(\widehat{\boldsymbol{\sigma}}\cdot {\bf g}_{12})\widehat{\boldsymbol
{\sigma}},
\quad {\bf v}_{2}^{\prime}={\bf v}_{2}+\mu_{rs}\left(
1+\alpha_{rs}^{-1}\right) (\widehat{\boldsymbol{\sigma}}\cdot {\bf
g}_{12})\widehat{\boldsymbol{\sigma}}\;,
\end{equation}
where ${\bf g}_{12}={\bf v}_1-{\bf v}_2$ is the relative velocity of the colliding pair,
$\widehat{\boldsymbol{\sigma}}$ is a unit vector directed along the centers of the two colliding spheres, and $\mu_{rs}=m_r/(m_r+m_s)$. The collision frequencies $\omega_{rs}$ can be seen as free parameters in the model. Their dependence on the coefficients of restitution $\alpha_{rs}$ can be chosen to optimize the agreement with  the results obtained from the Boltzmann equation  for IHS.

The relevant hydrodynamic fields in a mixture are the number densities $n_r$, the flow velocity  ${\bf u}$, and the granular temperature $T$. They are defined in terms of the distributions $f_r$ as
\begin{equation}
\label{2.2.1}
n_r=\int d{\bf v} f_r({\bf v}),
\end{equation}
\begin{equation}
\label{2.4}
 \rho{\bf u}=\sum_r\rho_r{\bf u}_r=\sum_r\int d{\bf v}m_r{\bf v}f_r({\bf v}),
\end{equation}
\begin{equation}
\label{2.5}
nT=p=\sum_rn_rT_r=\sum_r\frac{m_r}{d}\int d{\bf v}V^2f_r({\bf v}),
\end{equation}
where $\rho_r=m_rn_r$ is the mass density of species $r$, $n=\sum_r n_r$ is the total number density, $\rho=\sum_r\rho_r$ is the
total mass density, ${\bf V}={\bf v}-{\bf u}$ is the peculiar velocity, and $p$ is the hydrostatic pressure. Furthermore, the third equality of Eq.\ (\ref{2.5}) defines the kinetic temperatures $T_r$ of each species, which measure their mean kinetic energies. For inelastic systems, in general $T_r\neq T$ so that the energy equipartition theorem does not apply.

The collision operators conserve the particle number of each species and the
total momentum, but the total energy is not conserved:
\begin{equation}
\label{2.6.0}
\int d{\bf v}J_{rs}[{\bf v}|f_{r},f_{s}]=0 \;,
\end{equation}
\begin{equation}
\label{2.6.1}
\sum_{r,s}\int d{\bf v}m_{r}{\bf v}J_{rs}
[{\bf v}|f_{r},f_{s}]={\bf 0},
\end{equation}
\begin{equation}
\label{2.6.2}
\sum_{r,s}\int d{\bf v}\frac{1}{2}m_{r}V^{2}J_{rs}
[{\bf v}|f_{r},f_{s}]=-\frac{d}{2}nT\zeta.
\end{equation}
Here $\zeta$ is identified as the ``cooling rate'' due to inelastic
collisions among all species. At a kinetic level, it is also convenient to introduce
the ``cooling rates'' $\zeta_r$ for the partial temperatures $T_r$. They are defined as
\begin{equation}
\label{2.7}
\zeta_r=\sum_s \zeta_{rs}=-\sum_s \frac{1}{dn_rT_r}\int d{\bf v}m_rV^{2}J_{rs}[{\bf v}|f_{r},f_{s}],
\end{equation}
where the second equality defines the quantities $\zeta_{rs}$. The total cooling rate $\zeta$ can be written in terms of the partial cooling rates $\zeta_r$ as
\begin{equation}
\label{2.7.1}
\zeta=\sum_r\, x_r\gamma_r\zeta_r,
\end{equation}
where $x_r=n_r/n$ is the mole fraction of species $r$ and $\gamma_r\equiv T_r/T$.

The macroscopic balance equations for the mixture follow from the conditions (\ref{2.6.0})--(\ref{2.6.2}).
They are given by
\begin{equation}
D_{t}n_{r}+n_{r}\nabla \cdot {\bf u}+\frac{\nabla \cdot {\bf j}_{r}}{m_{r}}
=0\;,  \label{2.9}
\end{equation}
\begin{equation}
D_{t}{\bf u}+\rho ^{-1}\nabla \cdot {\sf P}={\bf 0}\;,  \label{2.10}
\end{equation}
\begin{equation}
D_{t}T-\frac{T}{n}\sum_{r}\frac{\nabla \cdot {\bf j}_{r}}{m_{r}}+\frac{2}{dn}
\left( \nabla \cdot {\bf q}+{\sf P}:\nabla {\bf u}\right)
=-\zeta T\;. \label{2.11}
\end{equation}
In the above equations, $D_{t}=\partial _{t}+{\bf u}\cdot \nabla $ is the
material derivative,
\begin{equation}
{\bf j}_{r}=m_{r}\int d{\bf v}\,{\bf V}\,f_{r}({\bf v})
\label{2.12}
\end{equation}
is the mass flux for species $r$ relative to the local flow,
\begin{equation}
{\sf P}=\sum_{r}\,\int d{\bf v}\,m_{r}{\bf V}{\bf V}\,f_{r}({\bf  v})
\label{2.13}
\end{equation}
is the total pressure tensor, and
\begin{equation}
{\bf q}=\sum_{r}\,\int d{\bf v}\,\frac{1}{2}m_{r}V^{2}{\bf V}
\,f_{r}({\bf v})
\label{2.14}
\end{equation}
is the total heat flux. The balance equations (\ref{2.9})--(\ref{2.11}) apply regardless of the details of the model considered for inelastic collisions. However, the influence of the collision model appears through the dependence of the cooling rate and the hydrodynamic fluxes on the coefficients of restitution.

As happens for elastic collisions \cite{GS03}, the main advantage of using IMM is
that a velocity moment of order $k$ of the Boltzmann collision operator only involves moments of order less than or equal to $k$ \cite{BC02}. This allows one to determine the Boltzmann collisional moments without the explicit knowledge of the velocity distribution functions.  The first few moments of the Boltzmann collision operator $J_{rs}[f_r,f_s]$ have been explicitly evaluated in Ref.\ \onlinecite{G03}. In particular, the cooling rates $\zeta_{rs}$ are given by
\begin{equation}
\label{2.15}
\zeta_{rs}=\frac{2\omega_{rs}}{d}\mu_{sr}(1+\alpha _{rs})\left[1-\frac{\mu_{sr}}{2}(1+\alpha_{rs})
\frac{\theta_r+\theta_s}{\theta_s}+\frac{\mu_{sr}(1+\alpha _{rs})-1}{d\rho_sp_r}
{\bf j}_r\cdot {\bf j}_s\right],
\end{equation}
where $p_r=n_rT_r$ and
\begin{equation}
\label{2.16}
\theta_r=\frac{m_r}{\gamma_r}\sum_{s}m_s^{-1}.
\end{equation}
Equation (\ref{2.15}) can be used to fix the parameters $\omega_{rs}$.  The most natural choice to optimize the agreement with the IHS results is to adjust the cooling rates $\zeta_{rs}$ for IMM, Eq.\ (\ref{2.15}), to be the same as the ones found for IHS. Given that the cooling rates are not exactly known for IHS, one can estimate them by considering their local equilibrium approximation \cite{GD99}. With this choice, the collision frequencies $\omega_{rs}$ are given by
\begin{equation}
\label{2.17}
\omega_{rs}=\frac{\Omega_d}{\sqrt{\pi}}n_s\sigma_{rs}^{d-1}
\left(\frac{\theta_r+\theta_s}{\theta_r\theta_s}\right)^{1/2} v_0,
\end{equation}
where $\sigma_{rs}=(\sigma_r+\sigma_s)/2$, $\sigma_r$ being the diameter of particles of species $r$. In addition,  $v_0=(2T\sum_r m_r^{-1})^{1/2}$ is a thermal velocity defined in terms of the global temperature $T$. Upon deriving (\ref{2.17}) use has been made of the fact that the mass flux ${\bf j}_r$ vanishes in the local equilibrium approximation. In the remainder of this paper, we will take the choice (\ref{2.17}) for $\omega_{rs}$.

\section{Navier-Stokes hydrodynamic equations}
\label{sec3}

Needless to say, the usefulness of the balance equations (\ref{2.9})--(\ref{2.11}) is limited without further specification of the fluxes and the cooling rate on space and time. However, for sufficiently large space and time scales, one expects that the system achieves a hydrodynamic regime in which all the space and time dependence of the distribution function occurs through a functional dependence on the hydrodynamic fields. This functional dependence is made local in space and time by writing $f_r({\bf v})$ as a series expansion in powers of the gradients of the hydrodynamic fields. This special solution is called a {\em normal} solution and can be obtained by applying the Chapman-Enskog method \cite{CC70} to the BE. One important difference with respect to the conventional Chapman-Enskog method for ordinary gases is that the reference state (zeroth-order approximation) of the expansion is not the local equilibrium state but the so-called (local) homogeneous cooling state, whose explicit form is not known \cite{GD99}. Very recently, the Chapman-Enskog solution to the BE  (\ref{2.1}) has been worked out to first order in gradients of the fields. The corresponding constitutive equations found up to this order for the binary mixture are \cite{GA04}
\begin{equation}
\label{3.1}
{\bf j}_1=-\frac{m_1m_2n}{\rho}D\nabla x_1-\frac{\rho}{p}D_p\nabla p-
\frac{\rho}{T}D'\nabla T,\quad {\bf j}_2=-{\bf j}_1,
\end{equation}
\begin{equation}
\label{3.2}
{\bf q}=-T^2D''\nabla x_1-\kappa\nabla p-\lambda\nabla T,
\end{equation}
\begin{equation}
\label{3.3}
P_{k\ell}=p\delta_{k\ell}-\eta\left(\nabla_\ell u_k+
\nabla_k u_\ell-\frac{2}{d}\delta_{k\ell}\nabla \cdot {\bf u}\right).
\end{equation}
The transport coefficients are the diffusion coefficient $D$, the thermal diffusion coefficient $D'$, the pressure diffusion coefficient $D_p$, the Dufour coefficient $D''$, the thermal conductivity $\lambda$, the pressure energy coefficient $\kappa$ and the shear viscosity coefficient $\eta$. In contrast to the results previously derived for IHS \cite{GD02}, all the above transport coefficients have been {\em exactly} obtained in terms of the coefficients of restitution and the parameters of the mixture (masses, sizes, and composition). Technical details of the Chapman-Enskog solution to the BE (\ref{2.1}) can be found in Ref.\ \cite{GA04}. Here, for the sake of completeness, the explicit expressions of the Navier-Stokes transport coefficients as well as the cooling rate $\zeta$ are displayed in the Appendix \label{appA}.

The expressions for the mass flux (\ref{3.1}), the heat flux (\ref{3.2}), the pressure tensor (\ref{3.3}), and the cooling rate (\ref{a7}) provide the necessary constitutive equations to convert the balance equations (\ref{2.9})--(\ref{2.11}) into a closed set of six independent equations for the hydrodynamic fields. Since the irreversible fluxes have been represented in terms of the gradients of the mole fraction $x_{1}$, the pressure $p$, the temperature $T$, and the flow velocity ${\bf u}$, it is convenient to use these as the independent hydrodynamic variables. This means that, apart from the balance equations (\ref{2.10}) and (\ref{2.11}) for ${\bf u}$ and $T$, respectively, we also need the corresponding balance equations for $x_{1}$ and $p$. These equations  can be easily obtained from (\ref{2.9}) and (\ref{2.11}) and are given by
\begin{equation}
D_{t}x_{1}+\frac{\rho }{n^{2}m_{1}m_{2}}\nabla \cdot {\bf j}_{1}=0\;,
\label{3.5}
\end{equation}
\begin{equation}
D_{t}p+p\nabla \cdot {\bf u}+\frac{2}{nd}\left( \nabla \cdot {\bf q}+
{\sf P}:\nabla {\bf u}\right) =-\zeta p.  \label{3.6}
\end{equation}
Therefore, when the expressions of the fluxes and the cooling rate $\zeta$ are
substituted into the balance equations (\ref{2.10}), (\ref{2.11}), (\ref{3.5}),
and (\ref{3.6}) one gets a closed set of hydrodynamic equations for $
x_{1}$, ${\bf u}$, $T$, and $p$. These are the  Navier-Stokes equations for the granular binary mixture. They are
given by
\begin{equation}
D_{t}x_{1}=\frac{\rho }{n^{2}m_{1}m_{2}}\nabla \cdot \left( \frac{m_{1}m_{2}n
}{\rho }D\nabla x_{1}+\frac{\rho }{p}D_{p}\nabla p+\frac{\rho }{T}D'
\nabla T\right) \;,  \label{3.7}
\end{equation}
\begin{eqnarray}
\left( D_{t}+\zeta \right) p+\frac{d+2}{d}p\nabla \cdot {\bf u} &=&\frac{2}{
d}\nabla \cdot \left( T^{2}D^{\prime \prime }\nabla x_{1}+\kappa\nabla p+\lambda
\nabla T\right)  \nonumber \\
&&+\frac{2}{d}\eta \left( \nabla _{\ell }u_{k}+\nabla _{k}u_{\ell }-\frac{2}{
d}\delta _{k\ell }\nabla \cdot {\bf u}\right) \nabla _{\ell }u_{k},
\label{3.8}
\end{eqnarray}
\begin{eqnarray}
\left( D_{t}+\zeta \right) T+\frac{2}{dn}p\nabla \cdot {\bf u} &=&
-\frac{T}{n}\frac{m_2-m_1}{m_1m_2} \nabla \cdot
\left( \frac{m_{1}m_{2}n}{\rho }D\nabla x_{1}+\frac{\rho }{p}D_{p}\nabla p+
\frac{\rho }{T}D^{\prime }\nabla T\right)  \nonumber \\
&&+\frac{2}{d n}\nabla \cdot \left( T^{2}D^{\prime \prime }\nabla
x_{1}+\kappa\nabla p+\lambda \nabla T\right)  \nonumber \\
&&+\frac{2}{d n}\eta \left( \nabla _{\ell }u_{k}+\nabla _{k}u_{\ell }-\frac{2
}{d}\delta _{k\ell }\nabla \cdot \mathbf{u}\right) \nabla _{\ell }u_{k},
\label{3.9}
\end{eqnarray}
\begin{equation}
D_{t}u_{\ell}+\rho^{-1}\nabla_{\ell}p=\rho ^{-1}\nabla _{k}\eta \left( \nabla _{\ell
}u_{k}+\nabla _{k}u_{\ell }-\frac{2}{d}\delta _{k\ell }\nabla \cdot {\bf u}\right) \;.
\label{3.10}
\end{equation}

\section{Linear stability analysis of the hydrodynamic equations}
\label{sec4}

As said in the Introduction, one of the simplest application of the hydrodynamic equations is a stability analysis of the nonlinear hydrodynamic equations (\ref{3.7})--(\ref{3.10}) with respect to the homogeneous state for small initial perturbations. The linearization about the homogeneous solution yields partial differential equations with coefficients that are independent of space but depend on time since the reference (homogeneous) state is cooling. As in the monocomponent case \cite{BDKS98}, this time dependence can be eliminated through a change in the time and space variables, and a scaling of the hydrodynamic fields.

Let us assume that the deviations $\delta y_{\alpha}({\bf r},t)=y_{\alpha}({\bf r},t)-y_{H \alpha}(t)$ are small. Here, $\delta y_{\alpha}({\bf r},t)$ denotes the deviation of $\{x_1, {\bf u}, T, p\}$ from their values in the homogeneous state, which is indicated by the subscript $H$. We introduce the following dimensionless space and time variables:
\begin{equation}
\label{4.1}
\tau=\int_{0}^{t} dt' \nu_{H}(t'),\quad {\bf r}'=\frac{\nu_{H}(t)}{v_{0H}(t)}{\bf r},
\end{equation}
where $\nu_H(t)=(\Omega_d/4\sqrt{\pi})n_H\sigma_{12}^2v_{0H}$ is an effective collision frequency and $v_{0H}=\sqrt{2T_H(m_1+m_2)/m_1m_2}$. Since $\{x_{1H}, {\bf u}_H, T_H, p_H\}$ are evaluated in the homogeneous cooling state, then
\begin{equation}
\label{4.2}
\partial_t x_{1H}=0, \quad {\bf u}_H={\bf 0}, \quad \partial_t \ln T_H=\partial_t \ln p_H=-\zeta_H.
\end{equation}
A set of Fourier transformed dimensionless variables are then introduced as
\begin{equation}
\label{4.4}
\rho_{{\bf k}}(\tau)=\frac{\delta x_{1{\bf k}}(\tau)}{x_{1H}}, \quad
{\bf w}_{{\bf k}}(\tau)=\frac{\delta {\bf u}_{{\bf k}}(\tau)}{v_{0H}(\tau)},\quad
\theta_{{\bf k}}(\tau)=\frac{\delta T_{{\bf k}}(\tau)}{T_{H}(\tau)}, \quad
\Pi_{{\bf k}}(\tau)=\frac{\delta p_{{\bf k}}(\tau)}{p_{H}(\tau)},
\end{equation}
where $\delta y_{{\bf k}\alpha}\equiv \{\delta x_{1{\bf k}}, \delta {\bf u}_{\bf k}, \delta T_{\bf k},
\delta p_{\bf k}\}$ is defined as
\begin{equation}
\label{4.3}
\delta y_{{\bf k}\alpha}({\bf k}, \tau)=\int d{\bf r}'\; e^{-i{\bf k}\cdot {\bf r}'}\delta y_{\alpha}({\bf r}',\tau).
\end{equation}
Note that here the wave vector ${\bf k}$ is dimensionless. In terms of the above variables, the transverse velocity components ${\bf w}_{\mathbf{k}\perp}={\bf w}_{\bf k}-({\bf w}_{\bf k}\cdot \widehat{{\bf k}})\widehat{{\bf k}}$ (orthogonal to the wave vector ${\bf k}$) decouple from the other four modes and hence can be obtained more easily. They obey the equation
\begin{equation}
\left( \frac{\partial }{\partial \tau }-\frac{\zeta ^{*}}{2}+\eta ^{*
}k^{2}\right) {\bf w}_{{\bf k}\perp}={\bf 0},  \label{4.5}
\end{equation}
where $\zeta ^{*}=\zeta _{H}/\nu _{H}$ is given by Eq.\ (\ref{a7}) and
\begin{equation}
\label{4.11}
\eta^*=\frac{\nu_H \eta}{\rho_H v_{0H}^2}.
\end{equation}
The solution for ${\bf w}_{{\bf k}\perp}({\bf k}, \tau)$ reads
\begin{equation}
\label{4.13}
{\bf w}_{{\bf k}\perp}({\bf k}, \tau)={\bf w}_{{\bf k}\perp}(0)\exp[s_{\perp}(k)\tau],
\end{equation}
where
\begin{equation}
\label{4.14}
s_{\perp}(k)=\frac{1}{2}\zeta^*-\eta^* k^2.
\end{equation}
This identifies two shear (transversal) modes. We see from Eq.\ (\ref{4.14}) that there exists a critical wave number $k_{\perp}^{\text{c}}$ given by
\begin{equation}
\label{4.15}
k_{\perp}^{\text{c}}=\left(\frac{\zeta^*}{2\eta^*}\right)^{1/2}.
\end{equation}
This critical value separates two regimes: shear modes with $k\geq k_{\perp}^{\text{c}}$ always decay while those with $k< k_{\perp}^{\text{c}}$ grow exponentially.

The remaining modes are called longitudinal modes. They correspond to the set $\{ \rho_{{\bf k}}, \theta_{{\bf k}}, \Pi_{{\bf k}} \}$ along with the longitudinal velocity component $w_{{\bf k}||}={\bf w}_{\bf k}\cdot \widehat{{\bf k}}$ (parallel to ${\bf k}$). These modes are the solutions of the linear equation
\begin{equation}
\frac{\partial \delta y_{{\bf k}\alpha }(\tau )}{\partial \tau }=\left(
M_{\alpha \beta }^{(0)}+ikM_{\alpha \beta }^{(1)}+k^{2}M_{\alpha \beta
}^{(2)}\right) \delta y_{{\bf k}\beta }(\tau ),
\label{4.6}
\end{equation}
where $\delta y_{{\bf k}\alpha }(\tau )$ denotes now the four variables $\left\{ \rho _{{\bf k}},\theta _{{\bf k}},\Pi _{{\bf k}},w_{{\bf k}||}\right\}$. The matrices in Eq.\ (\ref{4.6}) are given by
\begin{equation}
{\sf M}^{(0)}=\left(
\begin{array}{cccc}
0 & 0 & 0 & 0 \\
-x_{1}\left(\partial \zeta ^*/\partial x_{1}\right) _{T,p} &
\frac{1}{2}\zeta ^* & -\zeta ^{*} & 0 \\
-x_{1}\left(\partial \zeta ^*/\partial x_{1}\right) _{T,p} &
\frac{1}{2}\zeta ^{*} & -\zeta ^{*} & 0 \\
0 & 0 & 0 & \frac{1}{2}\zeta ^{*}
\end{array}
\right),   \label{4.7}
\end{equation}
\begin{equation}
{\sf M}^{(1)}=\left(
\begin{array}{cccc}
0 & 0 & 0 & 0 \\
0 & 0 & 0 & -\frac{2}{d} \\
0 & 0 & 0 & -\frac{d+2}{d}\\
0 & 0 & -\frac{1}{2}\frac{\mu/(1+\mu)}{x_{1}\mu +x_{2}} & 0
\end{array}
\right),   \label{4.8}
\end{equation}
\begin{equation}
{\sf M}^{(2)}=\left(
\begin{array}{cccc}
-D^{*} & -x_{1}^{-1}D'^{*} & -x_{1}^{-1}D_{p}^{*} & 0\\
-x_{1}\left( \frac{2}{d}D''^{*}-\frac{1-\mu }{x_{1}\mu+x_{2}}D^{*}\right)  & \frac{1-\mu}{x_{1}\mu+x_{2}}D'^{*}-\frac{
2}{d}\lambda ^{*} & -\frac{2}{d}\kappa^{*}+\frac{1-\mu}{x_{1}\mu+x_{2}}
D_{p}^{*} & 0 \\
-\frac{2}{d}x_{1}D''^{*} & -\frac{2}{d}\lambda ^{*} &
-\frac{2}{d}\kappa^{*} & 0 \\
0 & 0 & 0 & -\frac{2}{d}(d-1)\eta ^{*}
\end{array}
\right).   \label{4.9}
\end{equation}
In these equations, $\mu=m_1/m_2$, $x_r\equiv x_{rH}$,  and we have introduced the reduced Navier-Stokes transport coefficients
\begin{equation}
\label{4.10}
D^*=\frac{\nu_H D}{n_Hv_{0H}^2},\quad D_p^*=\frac{\rho_H^2\nu_H D_p}{m_1m_2n_H^2v_{0H}^2},\quad
D'^{*}=\frac{\rho_H^2\nu_H D'}{m_1m_2n_H^2v_{0H}^2},
\end{equation}
\begin{equation}
\label{4.12}
D''^{*}=\frac{\nu_H T_HD''}{n_Hv_{0H}^2},\quad \kappa^*=\frac{\nu_H \kappa}{v_{0H}^2},\quad
\lambda^{*}=\frac{\nu_H \lambda}{n_Hv_{0H}^2}.
\end{equation}

The longitudinal modes have the form $\exp[s_{n}(k)\tau]$ for $n=1,2,3,4$, where $s_n(k)$ are the eigenvalues of the matrix ${\sf M}={\sf M}^{(0)}+i k {\sf M}^{(1)}+k^2 {\sf M}^{(2)}$, namely, they are the solutions of the quartic equation
\begin{equation}
\label{4.16}
\det |{\sf M}-s\openone|=0.
\end{equation}
The solution to (\ref{4.16}) for arbitrary values of $k$ is quite complex. Interestingly, the expression of ${\sf M}$ through first order in $k$ does not depend on the assumed forms of the mass flux, the pressure tensor, and the heat flux. As a consequence, the corresponding eigenvalues can be considered as known exactly up through this order. They are found to be the eigenvalues of the matrix of ${\sf M}^{(0)}$:
\begin{equation}
\label{4.17}
s_{n}(0)=\left(0, 0, -\frac{1}{2}\zeta ^{*},\frac{1}{2}\zeta ^{*}\right).
\end{equation}
Hence, at asymptotically long wavelengths (${\bf k}=0$) the spectrum of the linearized hydrodynamic equations is comprised of a decaying mode at $-\zeta ^{*}/2$, a two-fold degenerate mode at $0$, and a three-fold degenerate unstable mode at $\zeta ^{*}/2$. For finite $k$, the longitudinal modes can be calculated for small $k$ by a perturbation expansion:
\begin{equation}
\label{4.17bis}
s_n(k)=s_n^{(0)}+k s_n^{(1)}+k^2s_n^{(2)}+\cdots,
\end{equation}
where $s_n^{(0)}$ is given by Eq.\ (\ref{4.17}). Through order $k^2$, the coefficients (for $n=1, 2, 3, 4$) are
\begin{equation}
\label{n1}
s_n^{(1)}=0,\quad n=1, 2, 3, 4,
\end{equation}
\begin{eqnarray}
\label{n2}
s_1^{(2)}&=&s_2^{(2)}=\left( \frac{\partial \ln \zeta ^*}{\partial x_{1}}\right) _{T,p}(D'^*+D_p^*)-\frac{1}{2\zeta^*}\frac{\mu}{(x_{1}\mu +x_{2})(1+\mu)}\nonumber\\
& & +\frac{1}{2}\frac{1-\mu}{x_{1}\mu +x_{2}}(2D'^*+D_p^*)-\frac{1}{2}D^*,
\end{eqnarray}
\begin{eqnarray}
\label{n3}
s_3^{(2)}&=&-\left[2\left( \frac{\partial \ln \zeta ^*}{\partial x_{1}}\right) _{T,p}+\frac{1-\mu}
{x_{1}\mu +x_{2}}\right](D'^*+D_p^*)\nonumber\\
& & +\frac{d+1}{d\zeta^*}\frac{\mu}{(x_{1}\mu +x_{2})(1+\mu)}-\frac{2}{d}(\lambda^*+
\kappa^*),
\end{eqnarray}
\begin{equation}
\label{n4}
s_4^{(2)}=-2\frac{d-1}{d}\eta^*-\frac{1}{d\zeta^*}\frac{\mu}{(x_{1}\mu +x_{2})(1+\mu)}.
\end{equation}
Since the Navier-Stokes order only applies through order $k^2$, the solutions (\ref{4.17}) and  (\ref{n1})--(\ref{n4}) are relevant to the same order.

\begin{figure}
\includegraphics[width=0.4 \columnwidth]{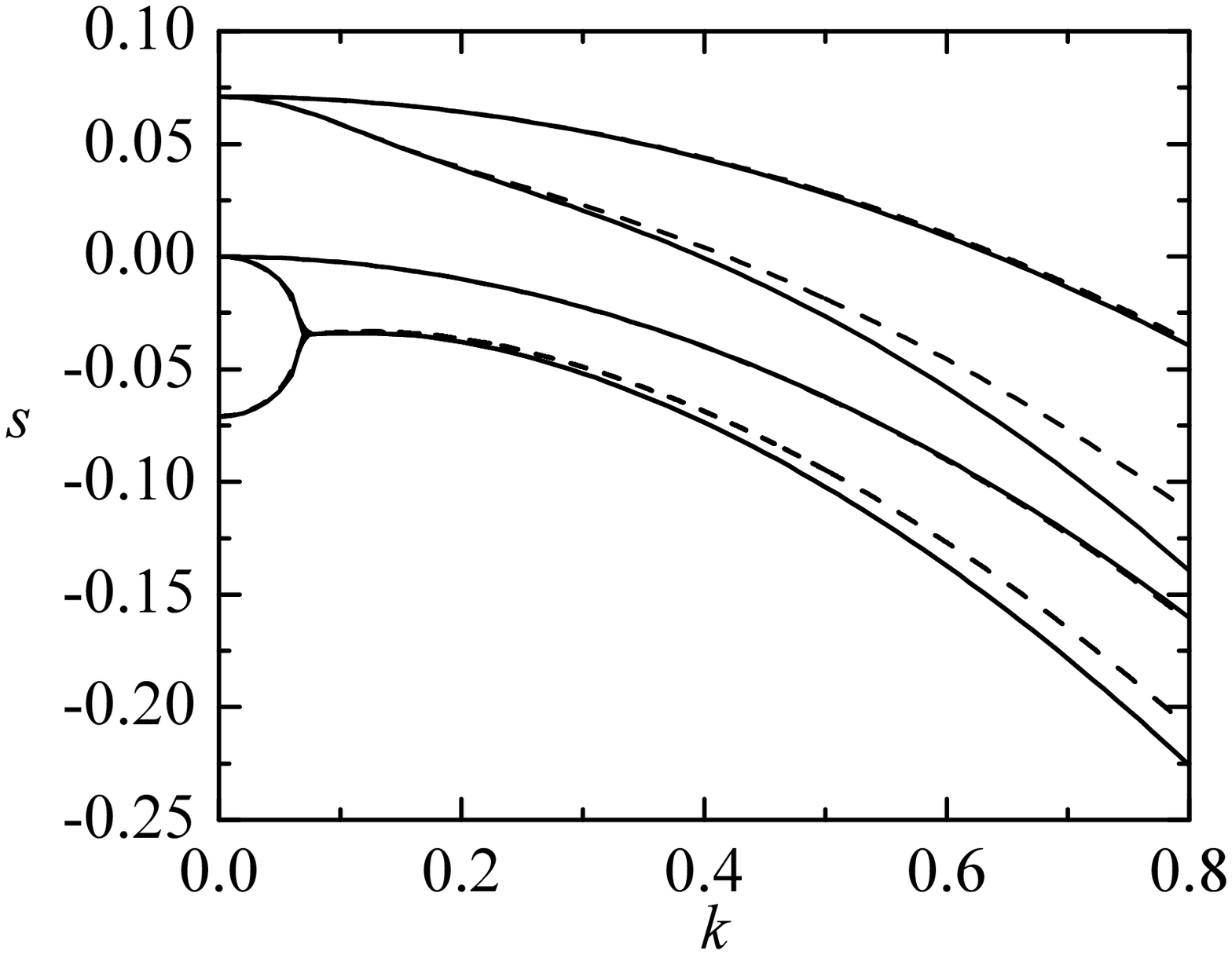}
\caption{Dispersion relations in the three-dimensional case for $\alpha_{11}=\alpha_{22}=\alpha_{12}\equiv \alpha=0.9$, $\sigma_1=\sigma_2$, $x_1=0.2$, and $m_1/m_2=4$. From top to bottom the curves correspond to the shear modes and the remaining four longitudinal modes. The solid lines are the results derived for IMM while the dashed lines refer to the results for IHS.
\label{fig1}}
\end{figure}

To illustrate the general dependence of the hydrodynamic modes on the reduced wavenumber $k$, in Fig.\ \ref{fig1} we show the dispersion relations for the case $\alpha_{11}=\alpha_{22}=\alpha_{12}\equiv \alpha=0.9$,  $\sigma_1/\sigma_2=1$, $x_1=0.2$, and $m_1/m_2=4$. From top to bottom the curves correspond to the shear modes, given by (\ref{4.14}), and the remaining four longitudinal modes. We only represent the real parts of the hydrodynamic modes. Furthermore, we have also included the results obtained in the case of  IHS by using the leading Sonine approximation for the transport coefficients \cite{GD02}. It is seen that the shear mode and one of the longitudinal modes (``heat'' mode) are positive for $k< k_{\perp}^{\text{c}}$ and $k< k_{h}^{\text{c}}$, respectively. Here, $k_{\perp}^{\text{c}}$ is given by Eq.\ (\ref{4.15}) while the critical value $k_{h}^{\text{c}}$ for the heat mode may be found from the condition $\det|{\sf M}|=0$. Therefore, initial long wavelength perturbations of the homogeneous cooling state that excite shear and heat modes will grow exponentially, representing an instability of the reference state. In addition, we also observe that the results for IMM compare quite well with those given for IHS showing again the reliability of IMM as a toy model to capture the main trends observed for granular fluids.

\section{Onsager's reciprocal relations}
\label{sec5}

As a second application of the explicit knowledge of the Navier-Stokes transport coefficients, in this Section we explore the validity of Onsager's reciprocal relations for inelastic Maxwell mixtures. In the case of elastic collisions, Onsager's relations \cite{GM84} establish symmetry properties between the Onsager phenomenological coefficients $L_{sr}$, $L_{sq}$, $L_{qq}$, and $L_{qs}$ associated with the mass and heat fluxes of a gas mixture. These coefficients are defined through the (linear) constitutive equations\cite{GM84}
\begin{equation}
\label{5.1}
{\bf j}_s=-\sum_{r=1}^2 L_{sr}\left(\frac{\nabla
\mu_r}{T}\right)_T-L_{sq}\frac{\nabla T}{T^2},
\end{equation}
\begin{eqnarray}
\label{5.2}
{\bf J}_q &\equiv& {\bf q}-\frac{d+2}{2}T\sum_{s=1}^2\frac{{\bf j}_s}{m_s}\nonumber\\
&=&-L_{qq}  \nabla T-\sum_{s=1}^2L_{qs}\left(\frac{\nabla \mu_s}{T}\right)_T.
\end{eqnarray}
Here, we have introduced the gradient of chemical potential per unit mass ($\mu_s)$  given by
\begin{equation}
\label{5.3}
\left(\frac{\nabla \mu_s}{T}\right)_T=\frac{1}{m_s}\nabla \ln (x_sp).
\end{equation}
 Onsager's relations state \cite{GM84}
\begin{equation}
\label{5.4}
L_{sr}=L_{rs},\quad L_{sq}=L_{qs}.
\end{equation}
Relations (\ref{5.4}) are a consequence of time reversal invariance of the equations of motion of the individual particles. Since the main feature of a granular gas is the {\em irreversible} loss of kinetic energy in collisions, one expects relations (\ref{5.4}) not to apply when $\alpha_{rs}\neq 1$. Now we want to assess the effect of dissipation on the violation of Onsager's theorem.

To identify the Onsager phenomenological coefficients from the constitutive equations (\ref{5.1}) and (\ref{5.2}), Eqs.\ (\ref{3.1}) and (\ref{3.2}) must be rewritten in terms of the gradients of chemical potential $\mu_r$ and temperature $T$. To do that, first note that Eq.\ (\ref{5.3}) leads to the relation
\begin{equation}
\label{5.5}
\nabla x_1=\frac{\rho_1\rho_2}{n\rho}\frac{(\nabla \mu_1)_T-(\nabla \mu_2)_T}{T}-\frac{n_1n_2}{n\rho}(m_2-m_1)\nabla \ln p,
\end{equation}
where use has been made of the identity $\nabla x_1=-\nabla x_2$. When Eq.\ (\ref{5.5}) is substituted into Eqs.\ (\ref{3.1}) and (\ref{3.2}), one gets the following expressions for the mass flux ${\bf j}_1$ and the heat flux ${\bf J}_q$ defined in Eq.\ (\ref{5.2}):
\begin{equation}
\label{5.6}
{\bf j}_1=-\frac{m_1m_2\rho_1\rho_2}{\rho^2}D\frac{(\nabla \mu_1)_T-(\nabla \mu_2)_T}{T}-C_p\nabla \ln p  -\frac{\rho}{T}D' \nabla T,
\end{equation}
\begin{eqnarray}
\label{5.7}
{\bf J}_q&=&
-\left(\frac{\rho_1\rho_2T^2}{n\rho}D''-\frac{d+2}{2}T\frac{m_2-m_1}{\rho^2}\rho_1\rho_2 D\right)
\frac{(\nabla \mu_1)_T-(\nabla \mu_2)_T}{T}\nonumber\\
& & -C_p'\nabla \ln p-\left(\lambda-\frac{d+2}{2}\frac{m_2-m_1}{m_1m_2}\rho D'\right)\nabla T,
\end{eqnarray}
where
\begin{equation}
\label{5.8}
C_p=\rho D_p-\frac{\rho_1\rho_2}{\rho^2}(m_2-m_1)D,
\end{equation}
and
\begin{equation}
\label{5.9}
C_p'= p \kappa-\frac{d+2}{2}T\frac{m_2-m_1}{m_1m_2}C_p-\frac{n_1n_2}{n\rho}T^2(m_2-m_1)D''.
\end{equation}

\begin{figure}
\includegraphics[width=0.4 \columnwidth]{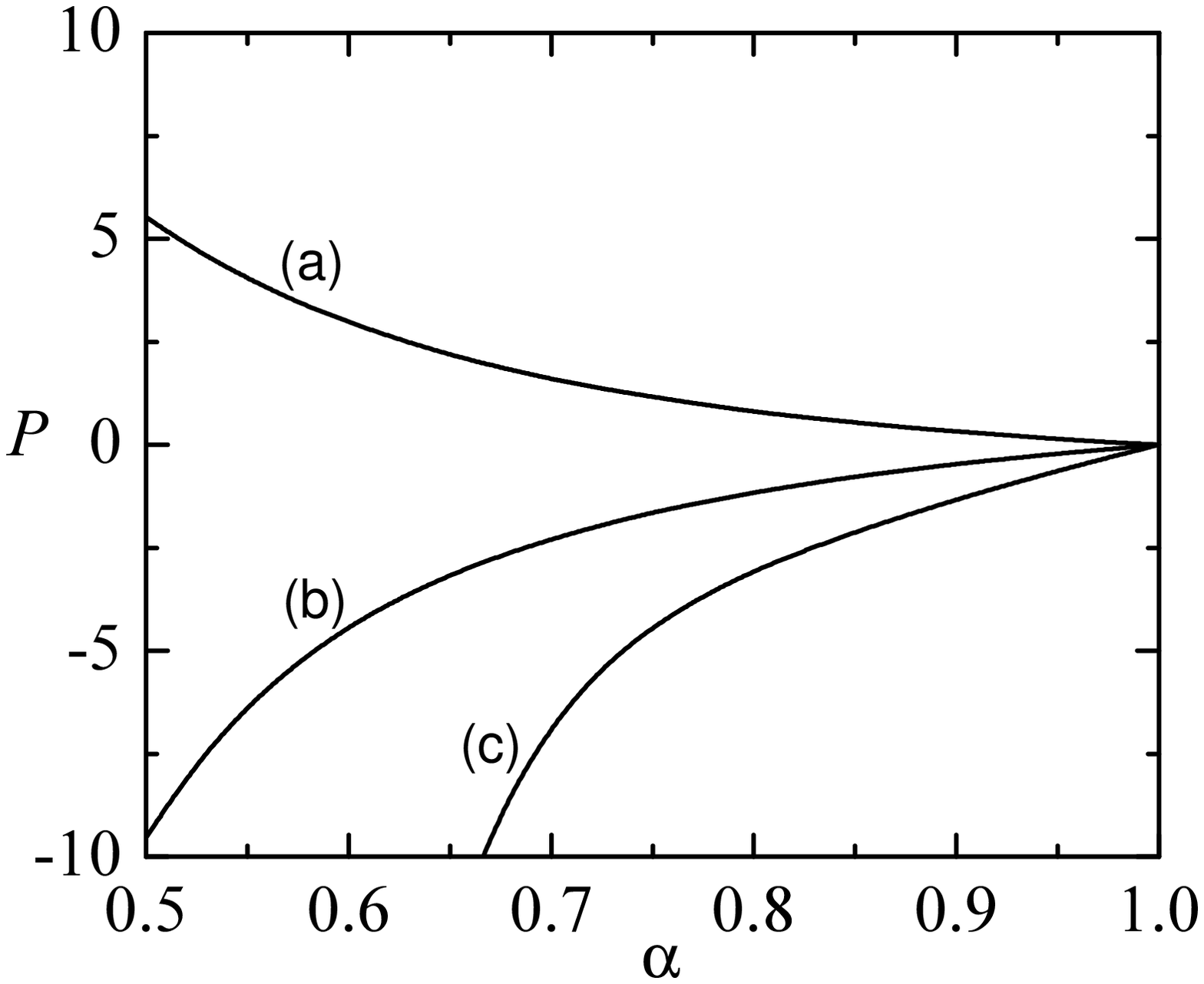}
\caption{Plot of the reduced function  $P(\alpha)$ as a function of the coefficient of restitution $\alpha$ in the three-dimensional case for $\sigma_1=\sigma_2$, $x_1=0.2$, and three different values of the mass ratio: (a) $m_1/m_2=0.5$, (b) $m_1/m_2=2$, and (c) $m_1/m_2=4$.
\label{fig2}}
\end{figure}

The Onsager phenomenological coefficients can be easily identified from Eqs.\ (\ref{5.6}) and (\ref{5.7}) when one takes into account the constitutive equations (\ref{5.1}) and (\ref{5.2}). They are given by
\begin{equation}
\label{5.10}
L_{11}= -L_{12}=\frac{m_1m_2\rho_1\rho_2}{\rho^2}D, \quad L_{1q}=\rho T D',
\end{equation}
\begin{equation}
\label{5.11}
L_{q1}= -L_{q2}=\frac{T^2\rho_1\rho_2}{n\rho}D''-\frac{d+2}{2}\frac{T \rho_1\rho_2}{\rho^2}(m_2-m_1) D, \quad L_{qq}=\lambda-\frac{d+2}{2}\rho \frac{m_2-m_1}{m_1m_2} D'.
\end{equation}
\begin{figure}
\includegraphics[width=0.4 \columnwidth]{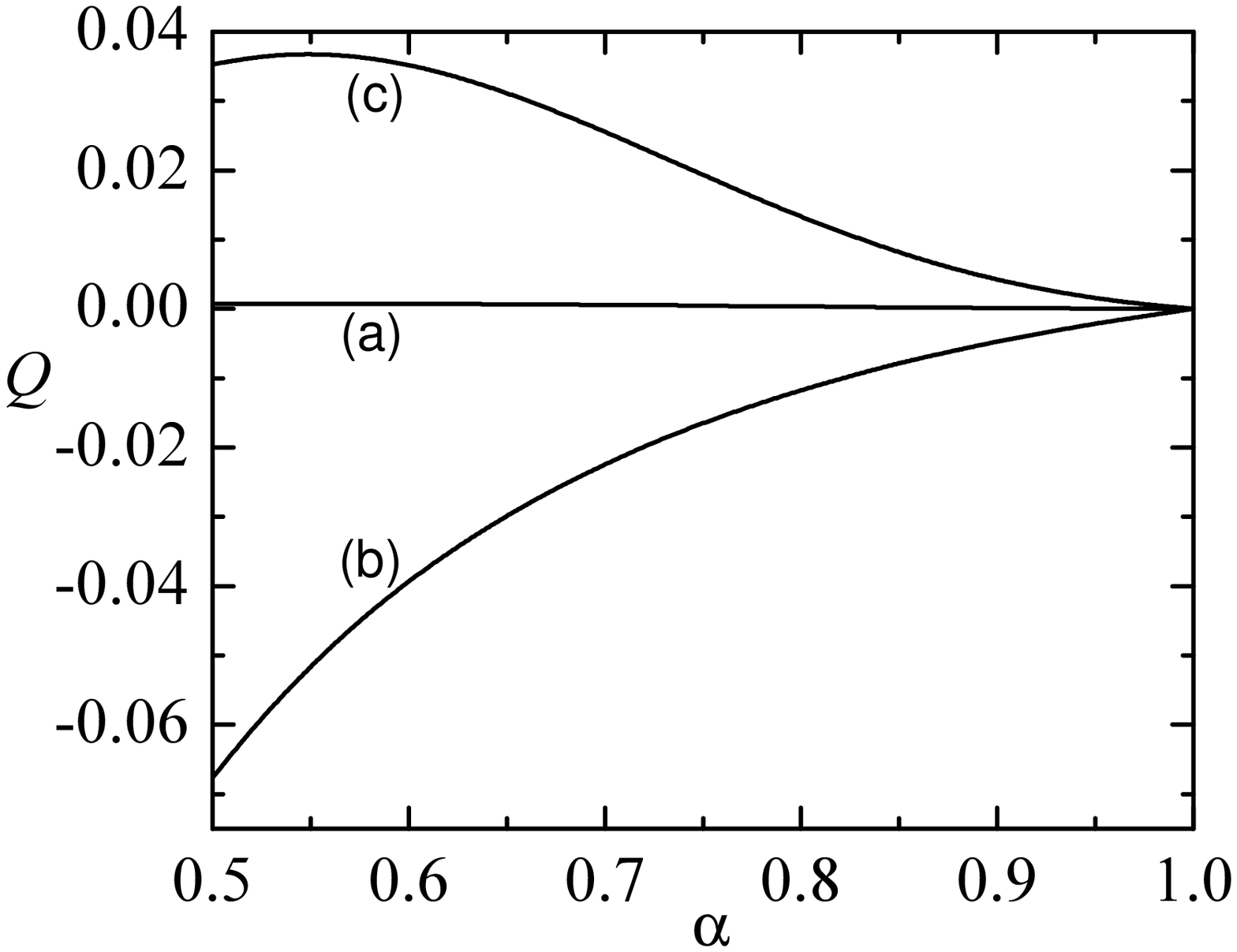}
\caption{Same as in Fig.\ \protect{\ref{fig2}}, but for the reduced function $Q(\alpha)$.
\label{fig3}}
\end{figure}
The first Onsager relation $L_{12}=L_{21}$ is trivially verified since, according to Eq.\ (\ref{a5}), the diffusion coefficient is symmetric under the exchange $1\leftrightarrow 2$. The second Onsager relation requires that $L_{1q}=L_{q1}$. In addition, according to Eqs.\ (\ref{5.6}) and (\ref{5.7}), there are two {\em new} contributions proportional to $\nabla p$ not present in the structure given by Eqs.\ (\ref{5.1}) and (\ref{5.2}) for the fluxes ${\bf j}_1$ and ${\bf J}_q$, respectively. Consequently, Onsager's relations would require that these contributions should vanish for any value of $\alpha_{rs}$, i.e., $C_p=C_p'=0$. In terms of the reduced coefficients $\{D^*, D_p^*, D'^*, D''^*, \kappa^*, \lambda^*\}$ defined in the Appendix, the conditions $L_{1q}=L_{q1}$, $C_p=0$, and $C_p'=0$ lead, respectively, to the following conditions
\begin{equation}
\label{5.11bis}
P(\{\alpha_{rs}\})=0, \quad Q(\{\alpha_{rs}\})=0, \quad R(\{\alpha_{rs}\})=0,
\end{equation}
where
\begin{equation}
\label{5.12}
P(\{\alpha_{rs}\})\equiv D''^*-\frac{d+2}{2}\frac{1-\mu^2}{\mu}D^*-\frac{1+\mu}{\mu}\frac{x_2+\mu x_1}{x_1x_2}D'^*,
\end{equation}
\begin{equation}
\label{5.13}
Q(\{\alpha_{rs}\})\equiv   D_p^*-x_1x_2\frac{1-\mu}{x_2+\mu x_1}D^*,
\end{equation}
\begin{equation}
\label{5.14}
R(\{\alpha_{rs}\}) \equiv \kappa^*-\frac{d+2}{2}\frac{1-\mu^2}{\mu}Q- x_1x_2\frac{1-\mu}{x_2+\mu x_1}D''^*.
\end{equation}
\begin{figure}
\includegraphics[width=0.4 \columnwidth]{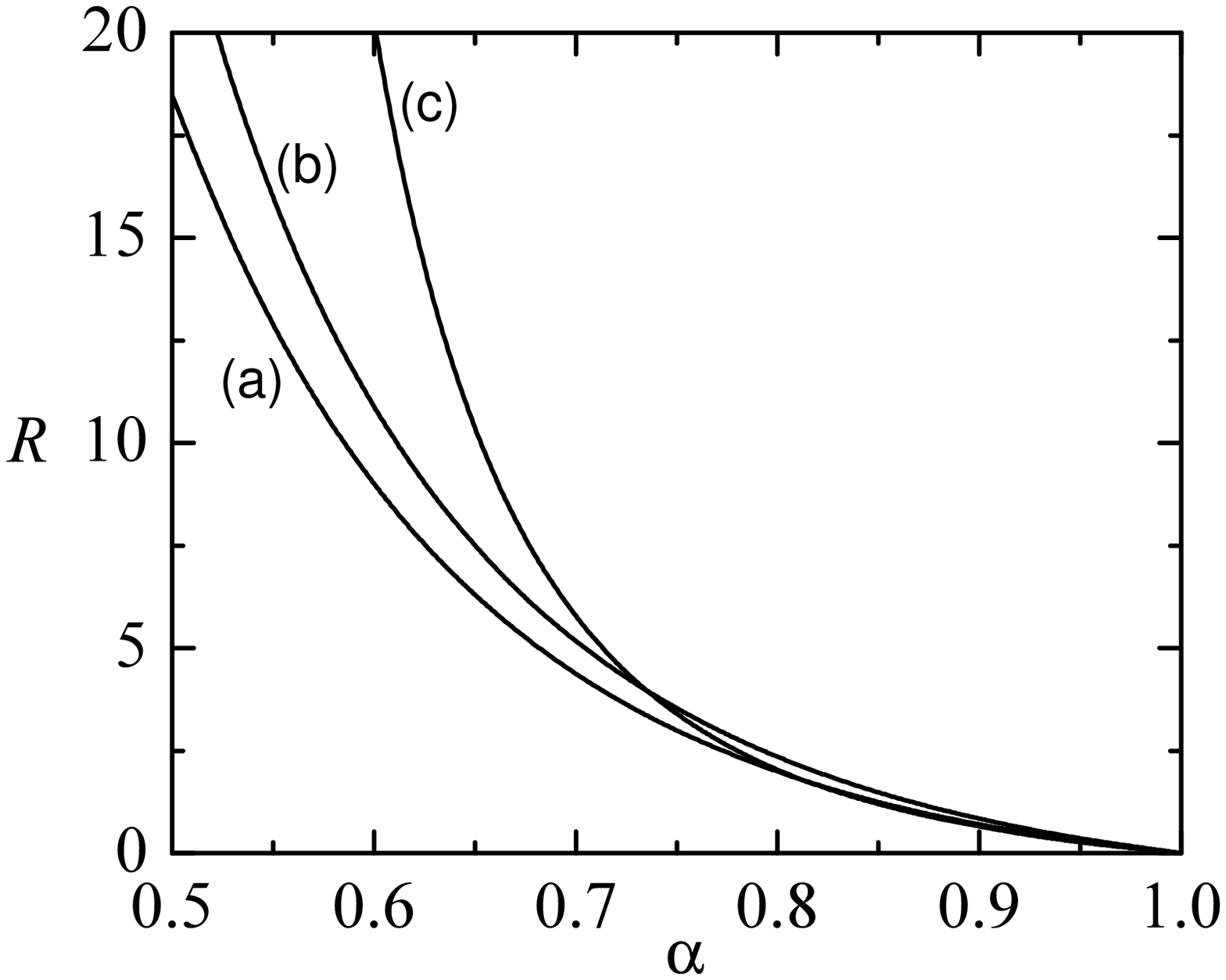}
\caption{Same as in Fig.\ \protect{\ref{fig3}}, but for the reduced function $R(\alpha)$.
\label{fig4}}
\end{figure}

In summary, if Onsager's relations held, then the reduced functions $P$, $Q$, and $R$ would vanish for any value of the parameters of the mixture. In the elastic limit ($\alpha_{rs}=1$), our results actually show that $P(1)=Q(1)=R(1)=0$, so that, Onsager's relations are exactly verified. Nevertheless, for inelastic collisions, the functions $P$, $Q$, and $R$ are different from zero, as expected. There are mainly three independent sources of discrepancy: (i) non-Gaussianity of the distribution functions of the homogeneous cooling state, which is measured through the cumulants $c_r$ and are given by Eq.\ (\ref{a18bis}); (ii) energy nonequipartition  as measured by the deviation of $\gamma$ from unity, and (iii) time evolution of the granular temperature, which is accounted for by the cooling rate $\zeta$. To illustrate the influence of dissipation on the breakdown of Onsager's symmetry relations, the functions $P$, $Q$, and $R$ are plotted in Figs.\ \ref{fig2}, \ref{fig3}, and \ref{fig4}, respectively, as functions of the coefficient of restitution $\alpha$. Here, for the sake of simplicity, we have assumed again a common coefficient of restitution ($\alpha_{rs}\equiv \alpha$) for $d=3$ with $x_1=0.2$, $\sigma_1=\sigma_2$, and different values of the mass ratio $m_1/m_2$. As might be expected, the deviation of $P$, $Q$, and $R$ from zero increases with decreasing $\alpha$, and in general is more significant as the mass disparity increases. In addition, while the magnitude of $Q(\alpha)$ is quite small, the function $R(\alpha)$ grows very fast with dissipation. The validity of the pure Onsager's relation $L_{1q}=L_{q1}$ is tested by the function $P(\alpha)$. Except for very weak dissipation, the violation of this reciprocal relation is quite important, especially in the case $m_1/m_2=4$.

\section{Concluding remarks}
\label{sec6}

In this paper we have derived the hydrodynamic equations for a granular binary mixture from the Boltzmann kinetic theory for {\em inelastic} Maxwell models (IMM). In the Boltzmann equation for IMM the collision rate of inelastic hard spheres (IHS) is replaced by an effective collision rate independent of the relative velocity of the two colliding particles. This simplification allows one to compute the velocity moments of the Boltzmann collision integrals without the explicit knowledge of the distribution functions. Thanks to this property, the Navier-Stokes transport coefficients of the mixture and the cooling rate have been {\em exactly} obtained in terms of dissipation and the parameters of the mixture (masses, sizes, composition) \cite{GA04}. These results contrast with the ones previously derived for IHS, where the transport coefficients were approximately determined by using the leading terms in a Sonine polynomial expansion of the distribution functions.

The explicit knowledge of the transport coefficients and the cooling rate allows one to study two different problems. As a first application, we have performed an analysis of the linearized hydrodynamic equations around the homogeneous cooling state. As in the monocomponent case for IHS \cite{BDKS98}, our stability analysis shows that the homogeneous cooling state is unstable to long enough wavelength perturbations and consequently becomes inhomogeneous for long times. Specifically, there are two shear modes and a longitudinal mode which become unstable for small values of the wavenumber. Therefore, small perturbations or fluctuations around the homogeneous state that excite the above modes will grow exponentially. As a second application of a hydrodynamic description, we have explored the breakdown of the Onsager relations between transport coefficients associated with the mass and heat fluxes of a granular mixture. For elastic systems, Onsager's reciprocity relations are a consequence of the fact that the mechanical equations of motion (classical as well as quantum mechanical) are symmetric with respect to time inversion. This microscopic property leads to Onsager's theorem. Given that the above time reversal invariance is broken in dissipative systems, violation of Onsager's theorem is expected for inelastic collisions. Here, we have analyzed the effect of the dissipation on such a violation. Our analysis shows that violation of Onsager's relations in granular gases has basically three distinct origins. First, the deviation of the homogeneous cooling state from the Gibbs state is responsible for the coefficients $c_1$ and $c_2$ [defined by Eq.\ (\ref{a18bis})] being different from zero. At a quantitative level, this effect is relatively quite small. Second, the effect of collisional cooling occurs through the presence of the cooling rate $\zeta$. Finally, the effect of different partial temperatures is expressed by the factors $\gamma_r$ and $\theta_r$ in the different terms involved in the calculation of the transport coefficients. Each one of these effects is a different reflection of dissipation in collisions.

In summary, there is growing theoretical support for the usefulness of a hydrodynamic description for granular systems under rapid flow conditions. However, in spite of the similarities between granular and normal fluids, the extension of properties of ordinary fluids to those with inelastic collisions must be carried out with caution. As shown here, the homogeneous (cooling) state is unstable to long wavelength perturbations and the familiar Onsager relations do not apply. These are some examples which makes these systems quite different from gases of elastic particles, like molecular gases.

\acknowledgments

We acknowledge the partial support of the Ministerio de Educaci\'on y Ciencia (Spain) through Grant No. FIS2004-01399 (partially financed by FEDER funds) in the case of V.G. and Grant No. ESP2003-02859 in the case of J.M.M.

\appendix
\section{Explicit expressions of the transport coefficients}
\label{appA}

In this Appendix we display the expressions of the transport coefficients $\{D, D_p, D', D'', \kappa, \lambda, \eta\}$ defining the mass flux (\ref{3.1}), the heat flux (\ref{3.2}), and the pressure tensor (\ref{3.3}). Let us consider first the coefficients associated with the mass flux. The expressions of the coefficients $D'$, $D_p$, and $D$ are given, respectively, by \cite{GA04}
\begin{equation}
\label{a3}
D'=-\frac{\zeta^*}{2\nu^*}D_p,
\end{equation}
\begin{equation}
\label{a4}
D_p=\frac{n_1T}{\rho\nu_0}\left(\gamma_1-\frac{\mu}{x_2+x_1\mu}\right)\left(\nu^*-\frac{3}{2}\zeta^*+
\frac{\zeta^{*2}}{2\nu^*}\right)^{-1},
\end{equation}
\begin{equation}
\label{a5}
D=\frac{\rho T}{m_1m_2\nu_0}\left[\left(\frac{\partial}{\partial x_1}x_1\gamma_1\right)_{p,T}+
\left(\frac{\partial \zeta^*}{\partial x_1}\right)_{p,T}\left(1-\frac{\zeta^*}{2\nu^*}
\right)\frac{\rho \nu_0}{p}D_p\right]\left(\nu^*-\frac{1}{2}\zeta^*\right)^{-1},
\end{equation}
where
\begin{equation}
\label{a2}
\nu_0=\frac{\Omega_d}{4\sqrt{\pi}}n\sigma_{12}^{d-1}v_0,\quad v_0=\sqrt{2T\frac{m_1+m_2}{m_1m_2}}.
\end{equation}
In these equations, $\mu=m_1/m_2$ is the mass ratio, $\zeta^*\equiv\zeta/\nu_0$, and we have introduced the (reduced) collision frequency
\begin{equation}
\label{a6}
\nu^*=\frac{4}{d}\left(x_2\mu_{21}+x_1\mu_{12}\right) \left(\frac{\theta_1+\theta_2}{\theta_1\theta_2}\right)^{1/2}
(1+\alpha_{12}),
\end{equation}
where $\theta_1=1/(\mu_{21}\gamma_1)$ and $\theta_2=1/(\mu_{12}\gamma_2)$. The temperature ratio $\gamma\equiv T_1/T_2$ is determined from the condition $\zeta_1=\zeta_2=\zeta$. In reduced units, the cooling rate $\zeta_1^*\equiv\zeta_1/\nu_0$ is given by
\begin{equation}
\label{a7}
\zeta_1^*=\frac{2\sqrt{2}}{d}x_1\left(\frac{\sigma_1}{\sigma_{12}}\right)^{d-1}
\theta_1^{-1/2}(1-\alpha_{11}^2)+\frac{8}{d}x_2\mu_{21}\left(\frac{\theta_1+\theta_2}{\theta_1\theta_2}\right)^{1/2}
(1+\alpha_{12})\left[1-\frac{\mu_{21}}{2}(1+\alpha_{12})\frac{\theta_1+\theta_2}{\theta_2}\right],
\end{equation}
while the expression of $\zeta_2^*\equiv\zeta_2/\nu_0$ can be easily obtained from (\ref{a7}) by the exchange $1\leftrightarrow 2$. It must be noted that $\zeta^*$ depends explicitly on $x_1$ and also implicitly through its dependence on the temperature ratio $\gamma$. Since ${\bf j}_{1}=-{\bf j}_{2}$ and $\nabla x_{1}=-\nabla x_{2}$, it is expected that  $D$ is symmetric with respect to exchange of particles $1$ and $2$ while $D_{p}$ and $D'$ are
antisymmetric. This can be easily verified by noting that $x_{1}\gamma_{1}+x_{2}\gamma_{2}=1$.

The case of the heat flux is more involved.  The transport coefficients $D''$, $\kappa$, and $\lambda$ are given by
\begin{equation}
\label{a7bis}
D''=D_1''+D_2'', \quad \kappa=\kappa_1+\kappa_2,\quad \lambda=\lambda_1+\lambda_2,
\end{equation}
where
\begin{equation}
\label{a8}
D_r''=\frac{n}{(m_1+m_2)\nu_0} D_r''^*,\quad \kappa_r=\frac{T}{(m_1+m_2) \nu_0} \kappa_r^*,\quad
\lambda_r=\frac{nT}{(m_1+m_2)\nu_0}\lambda_r^*.
\end{equation}
By using matrix notation, the coupled set of six equations for the unknowns
\begin{equation}
\label{a9}
\{D_1''^*, D_{2}''^*,\kappa_{1}^*, \kappa_{2}^*, \lambda_{1}^*, \lambda_{2}^*\}
\end{equation}
can be written as
\begin{equation}
\label{a10}
\Lambda_{\sigma \sigma'}X_{\sigma'}=Y_{\sigma}.
\end{equation}
Here, $X_{\sigma'}$ is the column matrix defined by the set
(\ref{a9})  and $\Lambda_{\sigma \sigma'}$ is the matrix
\begin{equation}
\label{a11}
\Lambda=\left(
\begin{array} {cccccc}
\nu_{11}^*-\frac{3}{2}\zeta^*& \nu_{12}^*&
-\left(\partial \zeta^*/\partial x_{1}\right)_{p,T}&0&
-\left(\partial \zeta^*/\partial x_{1}\right)_{p,T}&0 \\
\nu_{21}^*&\nu_{22}^*-\frac{3}{2}\zeta^*&0&
-\left(\partial \zeta^*/\partial x_{1}\right)_{p,T}&0&
-\left(\partial \zeta^*/\partial x_{1}\right)_{p,T}\\
0& 0& \nu_{11}^*-\frac{5}{2}\zeta^*& \nu_{12}^*&
-\zeta^*&0\\
0& 0 & \nu_{21}^* & \nu_{22}^*-\frac{5}{2}\zeta^* & 0 &
-\zeta^*\\
0& 0& \zeta^*/2&0&\nu_{11}^*-\zeta^*& \nu_{12}^*\\
0& 0& 0&\zeta^*/2&\nu_{21}^*&\nu_{22}^*-\zeta^*
\end{array}
\right),
\end{equation}
where
\begin{eqnarray}
\label{a20}
\nu_{11}^*&=&-\sqrt{2}x_1\left(\frac{\sigma_1}{\sigma_{12}}\right)^{d-1}\theta_1^{-1/2}
\frac{(1+\alpha_{11})}{d(d+2)}\left[\alpha_{11}(d+8)-5d-4\right]
\nonumber\\
& &
-4x_2\left(\frac{\theta_1+\theta_2}{\theta_1\theta_2}\right)^{1/2}\mu_{21}
\frac{(1+\alpha_{12})}{d(d+2)}
\left\{\mu_{21}(1+\alpha_{12})\left[d+8-3\mu_{21}(1+\alpha_{12})\right]-3(d+2)\right\},\nonumber\\
\end{eqnarray}
\begin{equation}
\label{a21}
\nu_{12}^*=-12 x_1\left(\frac{\theta_1+\theta_2}{\theta_1\theta_2}\right)^{1/2}
\mu_{12}\mu_{21}^2\frac{(1+\alpha_{12})^3}{d(d+2)}.
\end{equation}
The elements of the column matrix ${\bf Y}$ are
\begin{equation}
\label{a13}
Y_1=-\frac{m_1m_2\nu_0}{\rho T}D\Delta_{12}^*+\frac{d+2}{2}\mu_{12}^{-1}\frac{\partial}{\partial x_1}\left[\left(1+\frac{c_1}{2}\right)x_1\gamma_1^2\right]_{p,T},
\end{equation}
\begin{equation}
\label{a14}
Y_2=\frac{m_1m_2\nu_0}{\rho T}D\Delta_{21}^*+\frac{d+2}{2}\mu_{21}^{-1}\frac{\partial}{\partial x_1}\left[\left(1+\frac{c_2}{2}\right)x_2\gamma_2^2\right]_{p,T},
\end{equation}
\begin{equation}
\label{a15}
Y_3=-\frac{\rho \nu_0}{p} D_p\Delta_{12}^*+\frac{d+2}{2}\frac{x_1\gamma_1}{\mu_{12}}\left[\gamma_1\left(1+\frac{c_1}{2}\right)
-\frac{\mu}{x_2+\mu x_1}\right],
\end{equation}
\begin{equation}
\label{a16}
Y_4=\frac{\rho \nu_0}{p}D_p\Delta_{21}^*+\frac{d+2}{2}\frac{x_2\gamma_2}{\mu_{21}}\left[\gamma_2
\left(1+\frac{c_2}{2}\right)-\frac{1}{x_2+\mu x_1}\right],
\end{equation}
\begin{equation}
\label{a17}
Y_5=-\frac{\rho \nu_0}{p}D'\Delta_{12}^*+\frac{d+2}{2}\frac{x_1\gamma_1^2}{\mu_{12}}\left(1+\frac{c_1}{2}\right),
\end{equation}
\begin{equation}
\label{a18}
Y_6=\frac{\rho \nu_0}{p}D'\Delta_{21}^*+\frac{d+2}{2}\frac{x_2\gamma_2^2}{\mu_{21}}\left(1+\frac{c_2}{2}\right).
\end{equation}
Here, we have introduced the cumulants $c_r$ measuring the deviations of the zeroth-order distribution functions $f_r^{(0)}$ from a Maxwellian,
\begin{equation}
\label{a18bis}
c_r=\frac{2}{d(d+2)}\frac{m_r^2}{n_rT_r^2}\int d{\bf v} V^4 f_r^{(0)}-2.
\end{equation}
The reduced quantities $\Delta_{rs}^*$ are given by
\begin{eqnarray}
\label{a19}
\Delta_{12}^*&=&-\frac{1}{\sqrt{2}}x_1
\left(\frac{\sigma_1}{\sigma_{12}}\right)^{d-1}\frac{\gamma_1}{\mu_{12}}\theta_1^{-1/2}
\frac{(1+\alpha_{11})}{d(d+2)}\left[\alpha_{11}(d^2-2d-8)+3d(d+2)\right]
\nonumber\\
& &
-2x_2\left(\frac{\theta_1+\theta_2}{\theta_1\theta_2}\right)^{1/2}\mu_{21}\gamma_2
\frac{(1+\alpha_{12})^2}{d}\left[d-3\mu_{21}(1+\alpha_{12})+2\right]
\nonumber\\
& &
+2x_1\left(\frac{\theta_1+\theta_2}{\theta_1\theta_2}\right)^{1/2}\gamma_1\frac{(1+\alpha_{12})}{d}
\left[d+3\mu_{21}^2(1+\alpha_{12})^2-6\mu_{21}(1+\alpha_{12})+2\right].\nonumber\\
\end{eqnarray}
The corresponding expressions for $\Delta_{21}^*$, $\nu_{22}^*$, and $\nu_{21}^*$ can be inferred from Eqs.\ (\ref{a20}), (\ref{a21}), and (\ref{a19}) by exchanging $1\leftrightarrow 2$.

The solution to Eq.\ (\ref{a10}) is
\begin{equation}
\label{a22}
X_{\sigma}=\left(\Lambda^{-1}\right)_{\sigma \sigma'}Y_{\sigma'}.
\end{equation}
This relation provides an explicit expression for the coefficients
$D_{r}''^*$, $\kappa_{r}^*$, and $\lambda_{r}^*$ in terms of the coefficients of restitution
and the parameters of the mixture.

Finally, the shear viscosity coefficient $\eta$ can be written as
\begin{equation}
\label{a23}
\eta=\frac{nT}{\nu_0}\left(\eta_1^*+\eta_2^*\right),
\end{equation}
where the expression of the (reduced) partial contributions $\eta_r^*$ is
\begin{equation}
\label{a24}
\eta_1^*=2\frac{x_1\gamma_1(2\tau_{22}^*-\zeta^*)-2x_2\gamma_2\tau_{12}^*}{\zeta^{*2}-2\zeta^*
(\tau_{11}^*+\tau_{22}^*)+4(\tau_{11}^*\tau_{22}^*-\tau_{12}^*\tau_{21}^*)}.
\end{equation}
The quantities $\tau_{11}^*$ and $\tau_{12}^*$ are given by
\begin{eqnarray}
\label{a25}
\tau_{11}^*&=&4\sqrt{2}x_1\left(\frac{\sigma_1}{\sigma_{12}}\right)^{d-1}
\frac{(1+\alpha_{11})(d+1-\alpha_{11})}{d(d+2)}\nonumber\\
& & +8x_2\frac{\mu_{21}
(1+\alpha_{12})}{d}\left(\frac{\theta_1+\theta_2}{\theta_1\theta_2}\right)^{1/2}
\left[1-\frac{\mu_{21}(1+\alpha_{12})}{d+2}\right],
\end{eqnarray}
\begin{equation}
\label{a26}
\tau_{12}^*=-\frac{8x_2}{d(d+2)}\left(\frac{\theta_1+\theta_2}{\theta_1\theta_2}\right)^{1/2}
\frac{\rho_1}{\rho_2}\mu_{21}^2
(1+\alpha_{12})^2.
\end{equation}
A similar expression can be obtained for $\eta_2^*$ by just making the exchanges $1 \leftrightarrow 2$

\end{document}